\documentclass[secthm,seceqn,amsthm,ussrhead,reqno]{amsart}
\usepackage{amsmath,latexsym}
\usepackage[psamsfonts]{amssymb}
\usepackage{times}

\usepackage[mathcal]{euscript}

 \numberwithin{equation}{section}

\newcommand{\bbT}{\mathbb T}

\renewcommand{\epsilon}{\varepsilon}

\newcommand{\be}{\begin{equation}}
\newcommand{\ee}{\end{equation}}
\newcommand{\no}{\nonumber}

\newcommand{\C}{\mathbb{C}}

\newcommand{\R}{\mathbb{R}}
\renewcommand{\S}{\mathbb{S}}
\newcommand{\T}{\mathbb{T}}

\newcommand{\Z}{\mathbb{Z}}

\newcommand{\cB}{{\mathcal B}}

\newcommand{\cH}{{\mathcal H}}

\newcommand{\cU}{{\mathcal U}}

\renewcommand{\Im}{{\ensuremath{\mathrm{Im}}}}

\renewcommand{\det}{\mathop{\mathrm{det}}}

{\bf}{\it}
\newtheorem{theorem}{Theorem}[section]
\newtheorem{lemma}[theorem]{Lemma}

\newtheorem{corollary}[theorem]{Corollary}
\newtheorem{assumption}[theorem]{Assumption}
\newtheorem{definition}[theorem]{Definition}

\newtheorem{remark}[theorem]{Remark}

\date{\today}

\begin{document}
\title[On the spectrum of an Hamiltonian in Fock space]
{On the spectrum of an Hamiltonian in Fock space. \\Discrete
spectrum Asymptotics}

\author{Sergio  Albeverio$^{1,2,3}$, Saidakhmat  N. Lakaev$^{4}$,
  Tulkin H. Rasulov $^{5}$}

\address{ $^1$ Institut f\"{u}r Angewandte Mathematik,
Universit\"{a}t Bonn, Wegelerstr. 6, D-53115 Bonn\ (Germany)}

\address{
$^2$ \ SFB 611, \ Bonn, \ BiBoS, Bielefeld - Bonn;}
\address{
$^3$ \ CERFIM, Locarno and Acc.ARch,USI (Switzerland) E-mail
albeverio@uni.bonn.de}

\address{
{$^4$ Samarkand State University, University Boulevard 15, 703004,
Samarkand (Uzbekistan)} \ {E-mail: lakaev@yahoo.com }}

\address{
{$^5$ Samarkand State University, University Boulevard 15, 703004,
Samarkand (Uzbekistan)} \ {E-mail: rth@mail.ru }}
\begin{abstract}
A model operator $H$ associated with the energy operator of a system
describing three particles in interaction, without conservation of
the number of particles, is considered. The precise location and
structure of the essential spectrum of $H$ is described. The
existence of infinitely many eigenvalues below the bottom of the
essential spectrum of $H$ is proved for the case where an associated
generalized Friedrichs model has a resonance at the bottom of its
essential spectrum. An asymptotics for the number $N(z)$ of
eigenvalues below the bottom of the essential spectrum is also
established. The finiteness of eigenvalues of $H$ below the bottom
of the essential spectrum is proved if the associated generalized
Friedrichs model has an eigenvalue with energy at the bottom of its
essential spectrum.
\end{abstract}
\maketitle



\section{Introduction}

The main goal of the present paper is to give a thorough
mathematical treatment of the spectral properties for a model
operator $H$  with emphasis on the asymptotics for the number of
infinitely many eigenvalues (Efimov's effect case). The model
operator $H$ is associated with a system describing three particles
in interaction, without conservation of the number of particles.

The Efimov  effect is one of the remarkable results in the spectral
analysis for continuous three-particle  Schr\"{o}dinger  operators:
if none of the three two-particle Schr\"{o}dinger operators
(corresponding to the two-particle subsystems) has negative
eigenvalues, but at least two of them have a zero energy resonance,
then this three-particle Schr\"{o}dinger operator has an infinite
number of discrete eigenvalues, accumulating at zero.

Since its discovery by Efimov in 1970 \cite{Efi} many works have
been devoted to this subject. See, for example
\cite{AHW,AmNo,DFT,FaMe,OvSi,Sob,Tam91,Tam94,Yaf74}.

The main result obtained by  Sobolev \cite{Sob} (see also
\cite{Tam94}) is an asymptotics of the form $\cU_0|log|\lambda||$
for the number of eigenvalues on the left of $\lambda,\lambda<0$,
where the coefficient ${\cU}_0$ does not depend on the two-particle
potentials $ v_\alpha $ and is a positive function of the ratios
$m_1/m_2,m_2/m_3$ of the masses of the three-particles.

Recently  the existence of the Efimov effect for $N$-body quantum
systems with $N\geq 4$ has been proved by X.P. Wang in \cite{Wang}.

In fact in \cite{Wang} a lower bound on the number of eigenvalues of
the total (reduced) Hamiltonian of the form
$$C_0|log(E_0-\lambda)|$$ is given, when $\lambda$ tends to $E_0$,
where $C_0$ is a positive constant and $E_0$ is the bottom of the
essential spectrum.

 The kinematics of the quantum systems of describing three quasi-particles
on lattices  is rather exotic. For instance,
 due to the fact that   the discrete analogue of
the Laplacian (or its generalizations) is not rotationally
invariant, the Hamiltonian of a system  does not separate into two
parts, one relating to the center-of-mass motion and the other one
to the internal degrees of freedom. In particular, the Efimov effect
exists only for the zero value of the three-particle quasi-momentum
$K\in \T^3$ (see, e.g., \cite{ALM98,ALzMahp04,ALtmf03,Ltmf91,Lfa93,
LAfa99,Mat} for relevant discussions and
\cite{FIC,GrSc,KM,Mat,MS,Mog91,RSIII,Yaf00,Zh} for the general study
of the low-lying excitation spectrum for quantum systems on
lattices).

In statistical physics \cite{MiSp, MaMi}, solid-state physics
\cite{Mog91} and the theory of quantum fields \cite{Frie} some
important problems arise where the number of quasi-particles is
bounded, but not fixed. The study of systems describing $n$
particles in interaction, without conservation of the number of
particles is reduced to the investigation of the spectral properties
of self-adjoint operators acting in the {\it {cut subspace
${\cH}^{(n)}$}} of the Fock space, consisting of $r\le n$ particles
\cite{Frie,MiSp,Mog91,MiZh}.

The perturbation problem of an operator (a generalized Friedrichs
model), with point and continuous spectrum (which acts in
$\cH^{(2)}$) has played a considerable role in the  spectral
problems connected with the quantum theory of fields \cite{Frie}.

In the present paper we consider the perturbation problem, with two-
and three-particle essential and point  spectrum. Under some
smoothness assumptions on the parameters of a family of generalized
Friedrichs models $h(p),\,p\in\T^3=(-\pi,\pi]^3$ we obtain the
following results:

(i) We describe the location and structure of the essential spectrum
of $H$ via the spectrum of $h(p),\,p \in \T^3.$

(ii) We prove that the operator $H$ has infinitely many eigenvalues
below the bottom of the essential spectrum $\sigma_{ess}(H),$ if the
operator $h(0)$ has a resonance with energy at the bottom of its
essential spectrum.  Moreover, we establish the following asymptotic
formula for the number $N(z)$ of eigenvalues of $H$ lying below
$z<m=\inf \sigma_{ess}(H)$
\begin{equation*}
\lim\limits_{z \to m-0}\frac{N(z)}{|\log |z-m||}={\cU}_0
\,(0<{\cU}_0 <\infty).
\end{equation*}

(iii) We prove the finiteness of eigenvalues of $H$ below the bottom
of $\sigma_{ess}(H),$ if $h(0)$ has an eigenvalue with energy at the
bottom of its essential spectrum.

We remark that  the presence of a zero energy resonance for the
Schr\"{o}dinger operators is due to the two-particle interaction
operators $V$, in particular, the coupling constant (if $V$ has in
front of it a couling constant) (see, e.g.,
\cite{AGH,Ltmf92,rauch,Yaf74} ).

It is remarkable that for the generalized Friedrichs model $h(0)$
the presence of a resonance with energy at the bottom of its
essential spectrum (consequently the existence of infinitely many
eigenvalues of $H$) is due to the annihilation and creation
operators acting in the {\it symmetric Fock space}.

We notice that the assertion (iii) is surprising and similar
assertions have not yet been proved for the three-particle
Schr\"odinger operators on $\R^3$ or $\Z^3$.

We remark that  the operator $H$ has been considered before, but
only the existence of infinitely many eigenvalues below the bottom
of the essential spectrum of $H$ has been announced in \cite{LRfa03}
and only the location of the essential spectrum of $H$ has been
described in terms of zeroes of the Friedholm determinant in
\cite{LRmn03} in the case where the functions $u,v$ and $w$ were
analytic.

The organization of the present paper is as follows. Section 1 is an
introduction to the whole work. In Section 2 the model operator is
described as a bounded self-adjoint operator $H$ in ${\cH}^{(3)}$
and the main results of the present paper are formulated. Some
spectral properties of $h(p),p\in\T^3$ are studied in Section 3.
 In Section 4 the location and structure of the essential spectrum of $H$ is studied.
 In Section 5
we prove the Birman-Schwinger principle for the operator $H.$ In
Section 6 the finiteness of the  number of eigenvalues of the
operator $H$ is established. In Section 7 an asymptotic formula for
the number of eigenvalues is proved. Some technical material is
collected in Appendix $A.$

Throughout the present paper we adopt the following conventions:
Denote by  $\T^3$   the three-dimensional torus, the cube
$(-\pi,\pi]^3$ with appropriately  identified sides. The torus
$\T^3$ will always be considered as an abelian group with respect to
the addition and multiplication by real numbers regarded as
operations on $\R^3$ modulo $(2\pi \Z)^3$.

For each $\delta>0$ the notation $U_{\delta}(0) =\{p\in
{\bbT}^3:|p|<\delta \}$ stands for a $\delta$-neighborhood of the
origin.

Let  $\cB (\theta,\T^3)$ with $1/2<\theta<1$,  be the Banach spaces
of H\"older continuous functions on   ${\bbT}^3$ with exponent
$\theta$ obtained by the closure of the space of smooth (periodic)
functions $f$ on ${\bbT}^3$ with respect to the norm
\begin{equation*}
\|f\|_{\theta}=\sup_{t, \ell\in {\bbT}^3 \atop l\neq 0 }\bigg
[|f(t)|+|\ell|^{-\theta}|f(t+\ell)-f(t)|\bigg ].
\end{equation*}

The set of functions $f: {\bbT}^3\to \R$ having continuous
derivatives up to order $n$ inclusive will be denoted
$C^{(n)}({\bbT}^3).$ In particular $C^{(0)}({\bbT}^3)=C({\bbT}^3)$
by our convention that $f^{(0)}(x)=f(x).$

\section{The model operator and
statement of  results}

Let us introduce some notations used in this work. Let ${\C}={\C}^1$
be the field of complex numbers and let $L_2({\T}^3)$ be the Hilbert
space of square-integrable (complex) functions defined on $\T^3$ and
$L_2^s(({\T}^3)^2)$ be the Hilbert space of square-integrable
symmetric (complex) functions on $({ \T}^3)^2.$

Denote by ${\cH}^{(3)}$ the direct sum of spaces
${\cH}_0=\C^1,\,{\cH}_1= L_2({\T}^3)$ and
${\cH}_2=L_2^{s}(({\T}^3)^2),$ that is, ${\cH}^{(3)}= {\cH}_0 \oplus
{\cH}_1 \oplus {\cH}_2.$

Let $H$ be the operator in ${\cH}^{(3)}$ with the entries $H_{ij}:
{\cH}_j\to {\cH}_i, i,j=0,1,2:$
$$
(H_{00}f_0)_0=u_0f_0,\quad (H_{01}f_1)_0=\int\limits_{{\T}^3}
v(q')f_1(q')dq',\quad H_{02}=0,
$$$$ H_{10}=H^*_{01},\quad
(H_{11}f_1)_1(p)=u(p)f_1(p), \quad
(H_{12}f_2)_1(p)=\int\limits_{{\T}^3} v(q')f_2(p,q')dq',
$$
$$
H_{20}=0,\quad H_{21}=H^*_{12},\quad
(H_{22}f_2)_2(p,q)=w(p,q)f_2(p,q),
$$
where $H^*_{ij}:{\cH}_i\to {\cH}_j, (j=i+1,i=0,1)$ denotes the
adjoint operator to $H_{ij}.$

 Here $f_i \in \cH_i,i=0,1,2,$ $ u_0$ is a real number, $u$
is a real-valued essentially bounded function
 on ${\T}^3,$ $v$ is a  real-valued function belonging to $L_2(\T^3)$
 and $w$ is a real-valued essentially bounded symmetric function on $({\T}^3)^2.$

Under these assumptions the operator $H$ is bounded and
self-adjoint in ${\cH}$.

We remark that the operators $H_{10}$ and $H_{21}$ resp. $H_{01}$
and $H_{12}$ defined in the Fock space are called creation resp.
annihilation operators.

Throughout this paper we assume  the following additional
assumptions.

\begin{assumption}\label{hypoth1}

$(i)$ The symmetric function $w$ on $({\T }^3)^2$ is even with
respect to $(p,q),$ and has a unique non-degenerate minimum at the
point $(0,0)\in ({\T}^3)^2$ and all
third order partial derivatives of $w$ belong to $\cB (\theta,(\T^3)^2).$ \\
$(ii)$ There exist positive definite matrix $ W$ and real numbers
$l_1, l_2 (l_1>0,l_2\not=0)$ such that
$$
\left( \frac{\partial^2 w(0,0)}{\partial p_i \partial p_j}
\right)_{i,j=1}^3= l_1 W,\,\, \left( \frac{\partial^2
w(0,0)}{\partial p_i \partial q_j} \right)_{i,j=1}^3= l_2 W.
$$
\end{assumption}

\begin{assumption}\label{hypoth2} The function $u\in C^{(2)}(\T^3)$
is even on ${\T}^3,$ has a unique non-degenerate minimum at the
point $0\in {\T}^3$ and the function $v\in \cB (\theta,\T^3),$ with
$1/2<\theta< 1,$ is even.
\end{assumption}

\begin{remark}
In fact in the present paper we use only the condition $v^2\in \cB
(\theta,\T^3).$
\end{remark}

\begin{remark}
Note that if the function $w$ resp. $u$ is even and has a unique
minimum at a point $(p_0,p_0)\in ({\T}^3)^2$ resp. $p_1\in {\T}^3,$
then $p_0=0$ resp. $p_1=0.$ Therefore, without loss of generality we
assume that the function $w$ resp. $u$ has a unique minimum at the
point $(0,0)\in (\T^3)^2$ resp. $0\in \T^3.$
\end{remark}

Set
$$
m=\min_{p,q\in {\T}^3} w(p,q),\quad M=\max\limits_{p,q\in {\T}^3}
w(p,q)
$$
and
\begin{equation*}
\Lambda(p,z)=\int\limits_{{\T}^3} \frac{v^2(t)dt}{w(p,t)-z},\, p\in
\T^3,\, z\leq m.
\end{equation*}

\begin{assumption}\label{hypoth3} Assume that\\
(i) The function $\Lambda(\cdot,m)$ has a unique maximum at $p=0\in \T^3.$\\
ii) There exist positive numbers $\delta$ and $c$ such that for all
nonzero $p\in U_\delta(0)$ the inequality
 $$
 \Lambda(0,m)-\Lambda(p,m)>c p^2
 $$
 holds.
\end{assumption}

\begin{remark}
Let $\varepsilon$ be a real-analytic conditionally negative definite
function on ${\T}^3$ with a unique non-degenerate minimum at the
origin and the function $v\in \cB (\theta,\T^3)$ is even and
\begin{align}\label{u va w} u(p)= \varepsilon (p)+c,\, w(p,q)=\varepsilon (p)+
\varepsilon (p+q)+ \varepsilon (q)
\end{align}
for some real $c.$ Then Assumptions \ref{hypoth1}, \ref{hypoth2} and
\ref{hypoth3}  are fulfilled (see Lemma \ref{examp}).
\end{remark}

To formulate the main results of the paper we introduce a family of
generalized Friedrichs model $h(p),\,p\in \T^3$ which acts in
$\cH^{(2)}\equiv\cH_0 \oplus \cH_1$ with the entries
\begin{align}\label{h}
(h_{00}(p)f_0)_0=u(p)f_0,\,\,h_{01}=\frac{1}{\sqrt{2}}H_{01},\\
h_{10}=h^*_{01}, \,\, (h_{11}(p)f_1)_1(q)=w_p(q)f_1(q),\no
\end{align}
where $w_p(q)=w(p,q).$

Let the operator $h_0(p),\,p\in \T^3$ act in ${\cH}^{(2)}$ as
$$
h_0(p) \left( \begin{array}{ll}
f_0\\
f_1(q)
\end{array} \right)=
\left( \begin{array}{ll}
0\\
w_p(q)f_1(q)
\end{array} \right).
$$
The perturbation $h(p)-h_0(p)$ of the operator $h_0(p)$ is a
self-adjoint operator of rank 2. Therefore in accordance with the
invariance of the essential spectrum under finite rank perturbations
the essential spectrum $\sigma_{ess}(h(p))$ of $h(p)$ fills the
following interval on the real axis:
$$
\sigma_{ess}(h(p))=[m(p),M(p)],
$$
where the numbers $m(p)$ and $M(p)$ are defined by
\begin{equation*}
m(p)=\min_{q\in {\T}^3}w(p,q) \quad\mbox{and}\quad M(p)=\max_{q\in
{\T}^3} w(p,q).
\end{equation*}

\begin{remark}
We remark that for some $p\in \T^3$  the essential spectrum of
$h(p)$ may degenerate to the set consisting  of unique point
$[m(p),m(p)]$ and hence we can not state that the essential spectrum
of $h(p)$ is absolutely continuous for any $p\in \T^3.$ For example,
if the function $w$ is of the form \eqref{u va w}, where
\begin{align*}
\varepsilon (q)=3-cos q_1-cos q_2 -cos q_3,\quad q=(q_1,q_2,q_3) \in
{\T}^3
\end{align*}
and  $q=(\pi,\pi,\pi)\in \T^3$.
\end{remark}

The following theorem describes the essential spectrum of the
operator $H.$
\begin{theorem}\label{spec-1}
For the essential spectrum $\sigma_{ess}(H)$  of the operator $H$
the equality
$$
\sigma_{ess}(H)= \cup_{p\in {\T}^3} \sigma_d(h(p))\cup [m, M]
$$
holds, where $\sigma_d(h(p))$ is the discrete spectrum of the
operator $h(p),p\in\T^3$.
\end{theorem}

For any $p\in {\T}^3$ we define an analytic function (the Fredholm
determinant associated with the operator $h(p)$ ) $\Delta(p,z)$ in
${\C}\setminus [m(p),M(p)]$ by
\begin{equation}\label{Delta}
\Delta(p,z)=u(p)-z- \frac{1}{2} \int\limits_{{\T}^3}
\frac{v^2(q)dq}{w_p(q)-z}.
\end{equation}

Let $\sigma$  be the set of all complex numbers $z\in {\C}\setminus
[m(p),M(p)]$ such that the equality $\Delta(p,z)=0$  holds for some
$p\in {\T}^3.$

\begin{remark} We remark that in \cite{LRmn03}  the essential
spectrum of the operator $H$ has been described by means of zeroes
of the Fredholm determinant defined in \eqref{Delta} and by the
spectrum of multiplication operator $H_{22}$ as follows:
$$\sigma_{ess}(H)= \sigma \cup [m,M].$$

We note that the equality
\begin{equation*}
 \sigma=\cup_{p \in {\T}^3}\sigma_d(h(p))
\end{equation*}
holds (see Lemma \ref{ess.spec.H}).
\end{remark}

\begin{definition}
The set $\sigma$  resp. $[m,M]$ is called two- resp. three-particle
branch of the essential spectrum $\sigma_{ess}(H)$ of the operator
$H,$ which will be denoted by $\sigma_{two}(H)$ resp.
$\sigma_{three}(H).$
\end{definition}

The function $w_0(\cdot)$ has a unique non-degenerate minimum at
$q=0$ (see Lemma \ref{minimum}) and hence by Lebesgue's dominated
convergence theorem the finite limit
$$
\Delta(0,m)=\lim_{z\to m-0} \Delta(0,z)
$$
exists.

\begin{definition}\label{resonance0}
Let part $(i)$ of Assumption \ref{hypoth1} be fulfilled and
$u(0)\neq m.$ The operator $h(0)$ is said to have an $m$ energy
resonance if the number  $1$ is an eigenvalue of the operator
$$
(\mathrm{G}\psi)(q)=\frac{v(q)}{2(u(0)-m)} \int\limits_{{\T}^3}
\frac{v(t)\psi(t)dt}{w_0(t)-m},\,\,\psi \in {C(\T^3)}
$$
and the associated eigenfunction $\psi $ (up to a constant factor)
satisfies the condition $\psi(0)\neq 0.$
\end{definition}

\begin{remark}
Let part $(i)$ of Assumption \ref{hypoth1} be fulfilled and $v\in
\cB (\theta,\T^3).$\\
$(i)$ If $u(0)\leq m,$ then the equation $h(0)f=mf$ has only the
trivial solution $f\in {\C}^1\oplus L_1(\T^3).$ \\
$(ii)$ Assume that $u(0)>m$ and $\Delta(0,m)=0.$\\
a) If $v(0)\not=0,$ then the operator $h(0)$ has an $m$ energy
resonance and the vector $f=(f_0,f_1),$ where
\begin{equation}\label{f0f1}
f_0=const\neq 0,\,f_1(q)=-\frac{v(q)f_0}{\sqrt{2}(w_0(q)-m)},
\end{equation}
obeys the equation $ h(0)f=mf$ and hence $ f_1\in L_1(\T^3)\setminus L_2(\T^3)$
(see Lemma \ref{h0 resonans}).\\
b) If $v(0)=0,$ then the number $z=m$ is an eigenvalue of the
operator $h(0)$ and the vector $f=(f_0,f_1),$ where $f_0$ and
$f_1$ are defined by \eqref{f0f1}, obeys the equation $ h(0)f=mf$
and hence $ f_1\in L_2(\T^3)$ (see Lemma \ref{zeroeigen}).
\end{remark}

Let us denote by $\tau_{ess}(H)$ the bottom of the essential
spectrum $\sigma_{ess}(H)$ of the operator $H$ and by $N(z)$ the
number of eigenvalues of $H$ lying below $z \leq \tau_{ess}(H).$

The main result of this paper is the following

\begin{theorem}\label{fin} Let Assumptions \ref{hypoth1} and
\ref{hypoth2} be fulfilled.\\
(i) Assume that the number $z=m$ is an eigenvalue of $h(0)$ and let
Assumption \ref{hypoth3} be fulfilled. Then the operator $H$ has a
finite number of eigenvalues lying below $\tau_{ess}(H)=m.$\\
(ii)  Assume that the operator $h(0)$ has an $m$ energy resonance
and assume that part (i) of Assumption \ref{hypoth3} is fulfilled.
Then the operator $H$ has infinitely many eigenvalues lying below
$\tau_{ess}(H)=m$ and accumulating at $\tau_{ess}(H)=m.$ Moreover
the function $N(\cdot)$ obeys the relation
\begin{equation} \label{asym.K}
\lim\limits_{z \to m-0}\frac{N(z)}{|\log |z-m||}={\cU}_0
\,(0<{\cU}_0 <\infty).
\end{equation}
\end{theorem}
\begin{remark} The constant ${\cU}_0$ does not depend on the
function $v$  and is given as a positive function depending only on
the ratio $\frac {l_1}{l_2}$ (with $l_1,\,l_2$ as in Assumption
\ref{hypoth1}).
\end{remark}
\begin{remark}
We remark that if the conditions of Theorem \ref{fin} are fulfilled,
then   \\$\inf \sigma_{ess}(H)=m$ (see Lemma \ref{inf}).
\end{remark}
\begin{remark} We remark that in \cite{ALzMahp04} a result which is an
analogue of part (ii) of Theorem \ref{fin}, has been proven
 for the three-particle Schr\"odinger operators
associated with a system of three-particles on lattices interacting
by means zero-range pair potentials.
\end{remark}
\begin{remark} Clearly, the infinite cardinality of the
 discrete spectrum of $H$  lying on the l.h.s. of $m$
follows automatically from the positivity of ${\cU}_{0}.$
\end{remark}

\section{Spectral properties of the operators $h(p),p\in\T^3$}

In this section we study  some spectral properties of the family of
generalized Friedrichs model $h(p),\,\,p \in \T^3$ given by
\eqref{h}, which plays a crucial role in the  study of the spectral
properties of $H$. We notice that the spectrum and resonances of a
generalized Friedrichs model have been studied in detail in
\cite{ALaams97,LtsP86}.

\begin{lemma}\label{delta=0}
For any $p\in \T^3$ the operator $h(p)$ has an eigenvalue $z \in
{\C} \setminus [m(p),M(p)]$  if and only if $\Delta(p,z)=0.$
\end{lemma}
\begin{proof} If $u(p)\in \R \setminus
[m(p),M(p)]$ for any $p \in \T^3,$ then the equation
$h(p)f=m(p)f,\,f\in \cH^{(2)}$ has only trivial solution and hence
the value $u(p)\in \R \setminus [m(p),M(p)]$ can not be an
eigenvalue of the operator $h(p),$ where $\R$ is the set of real
numbers. Therefore, the number $z \in {\C} \setminus
([m(p),M(p)]\cup \{u(p)\})$ is an eigenvalue of the operator $h(p),p
\in \T^3$ if and only if (by the Birman-Schwinger principle) the
number $1$ is an eigenvalue of the operator
$$
(\mathrm{G}(p,z)\psi)(q)=\frac{v(q)}{2(u(p)-z)} \int\limits_{{\T}^3}
\frac{v(t)\psi(t)dt}{w_p(t)-z},\,\,\psi \in {L_2(\T^3)}.
$$

According to the Fredholm theorem  the number $\lambda=1$ is an
eigenvalue of the operator $G(p,z)$ if and only if $\Delta(p,z)=0.$
\end{proof}

\begin{lemma}\label{h0 resonans}
Let part $(i)$ of Assumption \ref{hypoth1}  be fulfilled. The
operator $h(0)$ has an $m$ energy resonance if and only if
$\Delta(0,m)=0$ and $v(0)\not=0.$
\end{lemma}
\begin{proof}
"Only If Part".  Suppose that the operator $h(0)$ has an $m$ energy
resonance. Then by Definition \ref{resonance0} the inequality
$u(0)\neq m$ holds and the equation
\begin{equation}\label{res-def}
\psi(q)= \frac{v(q)}{2(u(0)-m)} \int\limits_{{\T}^3}
\frac{v(t)\psi(t)dt}{w_0(t)-m},\,\,\psi\in {C(\T^3)}
 \end{equation}
 has a nontrivial solution $\psi \in C({\T^{3}})$ which satisfies the condition
 $\psi(0)\not=0.$

This solution is equal to the function $v$ (up to a constant factor)
and hence
$$ \Delta(0,m)= u(0)-m-\frac{1}{2}
\int\limits_{{\T}^3}\frac{v^2(t)dt}{w_0(t)-m} =0.
$$
"If Part". Let  the equality $ \Delta(0,m)= 0$  holds and $v(0)\neq
0$. Then the inequality $u(0)\neq m$ holds and the function $v \in
C({\T}^{3})$ is a solution of the equation \eqref{res-def},
  that is, by Definition \ref{resonance0}  the
operator $h(0)$ has an $m$ energy resonance.
\end{proof}

\begin{lemma}\label{zeroeigen} Let part $(i)$ of Assumption \ref{hypoth1}
 be fulfilled and $v\in \cB (\theta,\T^3).$ The number $z=m$ is an eigenvalue of the operator $h(0)$
 if and only if $\Delta(0,m)=0$ and  $v(0)=0.$
\end{lemma}
\begin{proof}
"Only If Part". Suppose $f=(f_0,f_1)$ is an eigenvector of the
operator $h(0)$ associated with the eigenvalue $z=m.$ Then $f_0$ and
$f_1$ satisfy the system of equations
\begin{equation}\label{sistema}
\left \lbrace
\begin{array}{llll}
(u(0)-m)f_0+ \frac{1}{\sqrt{2}}\int\limits_{{\T}^3} v(q')f_1(q')dq'=0\\
\frac{1}{\sqrt{2}}v(q)f_0+(w_0(q)-m)f_1(q)=0.
\end{array} \right.
\end{equation}
From \eqref{sistema} we find that $f_0$ and $f_1$ are given by
\eqref{f0f1} and from the first equation of (\ref{sistema}) we
derive the equality $ \Delta(0,m)=0. $

Since $w_0(\cdot)\in C^{(3)}(\T^3)$ and $v(\cdot)\in \cB
(\theta,\T^3)$ and the function $w_0(\cdot)$ has a unique
non-degenerate minimum at the origin we conclude that $f_1\in
L_2(\T^3)$ iff $v(0)=0.$

"If Part". Let $v(0)=0$ and  $\Delta(0,m)=0.$ Then the vector
$f=(f_0,f_1),$ where $f_0$ and $f_1$ are defined by \eqref{f0f1},
obeys the equation $h(0)f=mf$ and $f_1\in L_2(\T^3).$
\end{proof}

Let $v(\cdot)\in \cB (\theta,\T^3).$ For any $p,q\in \T^3$ and $z<m$
the inequality $w_p(q)-z>0$ implies the inequality
\begin{equation*}
\int\limits_{{\T}^3} \frac{v^2(t)dt}{w_p(t)-z}>0.
\end{equation*}

Since the function $w(\cdot,\cdot)$ has a unique non-degenerate
minimum at the point $(0,0)\in (\T^3)^2$  the integral
\begin{equation*}
\int\limits_{{\T}^3} \frac{v^2(t)dt}{w_p(t)-m}
\end{equation*}
is finite. The Lebesgue dominated convergence theorem yields the
equality
$$
\Delta(0,m)=\lim_{p\to 0} \Delta(p,m)
$$
and  hence the function $\Delta(\cdot,m)$ is continuous on $ \T^3.$

\begin{lemma}\label{neg eigen} Let part $(i)$ of Assumption \ref{hypoth1} be
fulfilled.\\
(i)  Assume that $\max\limits_{p\in\T^3}\Delta(p,m)<0.$ Then for any
$p
\in\T^3$ the operator $h(p)$ has a unique eigenvalue lying on the l.h.s. of $m.$\\
(ii) Assume that $\min\limits_{p\in\T^3}\Delta(p,m)<0$ and
$\max\limits_{p\in\T^3}\Delta(p,m)\ge 0.$ Then there exists a non
void open set $D\subset \T^3$ such that $D\neq \T^3$  and for any $p
\in D$ the operator $h(p)$ has a unique eigenvalue lying on the
l.h.s. of $m$ and for any $p \in \T^3\setminus D$  the operator
$h(p)$ has no
eigenvalues lying on the l.h.s. of $m.$\\
(iii) Assume that $\min\limits_{p\in\T^3}\Delta(p,m)\ge 0.$ Then for
any $p \in \T^3$ the operator $h(p)$ has no eigenvalues lying on the
l.h.s. of $m.$
\end{lemma}

\begin{proof} First we prove part $(ii).$

Let $\min\limits_{p\in\T^3}\Delta(p,m)<0$ and
$\max\limits_{p\in\T^3}\Delta(p,m)\ge 0.$

Introduce the notation
\begin{equation*}
D\equiv \{p\in \T^3: \Delta(p,m)<0\}.
\end{equation*}

Since $\T^3$ is compact and the function $\Delta(\cdot,m)$ is
continuous on $\T^3,$ there exist points $p_0,p_1\in\T^3$  such that
the inequalities
$$
\min\limits_{p\in\T^3}\Delta(p,m)=\Delta(p_0,m)<0\quad
\mbox{and}\quad \max\limits_{p\in\T^3}\Delta(p,m)=\Delta(p_1,m)\ge
0
$$
hold. Hence we have that $D$ is a non void open set and $D\neq
\T^3.$

For any $p\in\T^3$ the function $\Delta(p,\cdot)$ is continuous
and decreasing on $(-\infty,m]$ and
$$
\lim_{z\to -\infty} \Delta(p,z)=+\infty.
$$

Then for any $p\in D$  there exist a unique point $z(p)\in
(-\infty,m)$ such that $\Delta(p,z(p))=0.$ By Lemma \ref{delta=0}
for any $p\in D$ the point $z(p)$ is the unique eigenvalue of the
operator $h(p)$ lying on the l.h.s. of $m.$

For any $ p \in \T^3\setminus D$ and $z<m$ we have
\begin{equation*}
\Delta(p,z)>\Delta(p,m)\ge 0.
\end{equation*}

Hence by Lemma \ref{delta=0}  for each $p \in \T^3\setminus D$ the
operator $h(p)$ has no eigenvalue lying on the l.h.s. of $m.$

If $\max\limits_{p\in\T^3}\Delta(p,m)<0$ (resp.
$\min\limits_{p\in\T^3}\Delta(p,m)\ge 0$), then $D=\T^3$ (resp.
$D=\emptyset$) and the above analysis leads again to the case $(i)$
(resp. $(iii)$). The straightforward details are omitted.
\end{proof}

Set
$$
\C_+=\{z\in \C: Re z>0\},\quad  \R_+=\{x\in \R: x>0\},\quad
\R_+^0=\R_+\cup \{0\}.
$$

Let $w_0(\cdot,\cdot)$ be the function defined on
$U_{\delta}(0)\times \T^3,\,\delta>0$ sufficiently small, as
 \begin{equation}\label{w}
w_0(p,q)=w_p(q+q_0(p))-m(p),
\end{equation}
where $q_0(\cdot)\in C^{(3)}(U_{\delta}(0))$
 and  for any $p\in U_{\delta}(0)$ the
point $q_0(p)$  is the non-degenerate minimum of the function
$w_p(\cdot)$ (see Lemma \ref{minimum}). Here
$C^{(n)}(U_{\delta}(0))$ can be defined similarly to
$C^{(n)}(\T^3).$

For any $p\in U_\delta(0)$ we define an analytic function
$D(p,\zeta)$ in $\C_+$ by
\begin{equation*}
D(p,\zeta)=u(p)-m(p)+\zeta^2- \frac{1}{2} \int\limits_{{\T}^3}
\frac{v^2(q+q_0(p))dq}{w_0(p,q)+\zeta^2}.
\end{equation*}

\begin{lemma}\label{razlojeniya} Let Assumptions \ref{hypoth1}
and \ref{hypoth2} be fulfilled. Then there exist a number $\delta
>0$  such that

 i) For any  $\zeta\in \C_+$ the function
$D(\cdot,\zeta)$ is of class $C^{(2)}(U_{\delta}(0))$  and the
following decomposition
\begin{equation}
D(p,\zeta)=D(0,\zeta)+D^{res}(p,\zeta),
\end{equation}
holds, where $D^{res}(p,\zeta)=O(p^2)$ as $p\to 0$ uniformly in
$\zeta\in \R_+^0.$

ii)  The right-hand derivative of $D(0,\cdot)$ at $\zeta=0$ exists
and
$$
D(0,\zeta)=D(0,0)+2\sqrt{2}\pi^2  v^2(0) l_1^{-\frac{3}{2}}
  (det W)^{-\frac{1}{2}}\zeta+D^{res}(\zeta),
$$
where $D^{res}(\zeta)=O(\zeta^{1+\theta}),\, \zeta\in \R_+^0.$
\end{lemma}

\begin{remark}
An analogue lemma has been proven in \cite{ALzMahp04} in the case
where the functions $u(\cdot),\,b(\cdot)$ and $w(\cdot,\cdot)$ are
analytic on $\T^3$ and $(\T^3)^2,$ respectively.
\end{remark}

\begin{proof}
$i)$ Since $m(\cdot)\in C^{(3)}(U_{\delta}(0))$ by definition of the
function $D(\cdot,\cdot)$ and Assumptions \ref{hypoth1} and
\ref{hypoth2} we obtain that the function $D(\cdot,\zeta)$ is of
class $C^{(2)}(U_{\delta}(0))$ for any $\zeta\in \C_+.$

Using
 \begin{align*}
w_0(p,q)=\frac{l_1}{2}(Wq,q)+o(|p||q|^2)+o(|q|^2) \,\,as\,\,|p|,|q|
\to 0
\end{align*}
we obtain that there exists $C>0$ such that for any $\zeta\in
\R_+^0$ and $i,j=1,2,3$ the inequalities
\begin{align}\label{Estimate1}
&\Big |\frac{\partial^2}{\partial p_i \partial p_j}
 \frac{1 }{w_0(p,q)+\zeta^2}\Big |\leq \frac{C}{q^2},\,p,q\in U_\delta(0)
\end{align}
and
\begin{align}\label{Estimate2}
&\Big |\frac{\partial^2}{\partial p_i \partial p_j}
 \frac{1 }{w_0(p,q)+\zeta^2}\Big |\leq C,\,p\in U_\delta(0), q\in \T^3
 \setminus U_\delta(0)
\end{align}
hold.

The Lebesgue dominated convergence theorem implies that
$$
\frac{\partial^{2}}{\partial p_i\partial p_j}D(p,0)=\lim_{\zeta\to
0+}\frac{\partial^{2}}{\partial p_i\partial p_j}D(p,\zeta).
$$

Repeating the application of the Hadamard lemma (see \cite{Zor} V.1,
p. 512) we obtain
\begin{align*}
D(p,\zeta)=D(0,\zeta)+\sum_{i=1}^3\frac{\partial}{\partial p_i}
D(0,\zeta)p_i+\sum_{i,j=1}^{3}H_{ij}(p,\zeta) p_i p_j,
\end{align*}
where for any $\zeta\in \R_+^0$ the functions
$H_{ij}(\cdot,\zeta),\,i,j=1,2,3$ are continuous in $U_\delta(0)$
and
\begin{equation*}
    H_{ij}(p,\zeta)=\frac{1}{2}\int_{0}^1\int_{0}^1
    \frac{\partial^2}{\partial p_i \partial p_j}D(x_1x_2p,\zeta)dx_1dx_2.
\end{equation*}

The estimates \eqref{Estimate1} and \eqref{Estimate2} give
$$
|H_{i,j}(p,\zeta)|\leq \frac{1}{2}\int_{0}^1\int_{0}^1
    \Big |\frac{\partial^2}{\partial
p_i \partial p_j}D(x_1x_2p,\zeta)\Big |dx_1dx_2\leq
C\Big(1+\int\limits_{U_\delta(0)}\frac{v^2(q+q_0(p))dq}{q^2}\Big)
$$
for any $p\in U_\delta(0)$ uniformly in $\zeta\in \R_+^0.$

Since for any $\zeta\in \C_+$ the function $D(\cdot,\zeta)$ is even
in $U_\delta(0)$ we have
$$
 \left[\frac{\partial}{\partial p_i}
D(p,\zeta)\right]_{p=0}=0,\quad i=1,2,3.
$$

$ii)$ Now we prove that there exists the right-hand derivative of
$D(0,\cdot)$ at $\zeta=0$ and the following inequalities
\begin{align}\label{D-D}
&|D(0,\zeta)-D(0,0)|\leq C \zeta,\quad \zeta\in \R^0_+,
\end{align}
\begin{equation}\label{partial D}
\big |  \frac{\partial }{\partial \zeta}D(0,\zeta)- \frac{\partial
}{\partial \zeta}D(0,0) \big |<C \zeta^\theta ,\quad \zeta\in \R^0_+
\end{equation}
hold for some positive $C.$

Indeed, the function $D(0,\cdot)$ can be represented as
\begin{equation*}
D( 0, \zeta) =D_1(\zeta)+D_2(\zeta)
\end{equation*}
with
\begin{equation}\label{I1}
D_1(\zeta)=-\frac{1}{2}\int\limits_{U_{\delta}(0)}
\frac{v^2(q)}{w_0(0,q)+\zeta^2 }dq,\,\zeta\in \C_+
\end{equation}
and
$$
D_2(\zeta)=u(0)-m+\zeta^2-\frac{1}{2}\int\limits_{{\T}^3\setminus
U_{\delta}(0)} \frac{v^2(q)}{w_0(0,q)+\zeta^2 }dq,\,\zeta\in \C_+.
$$

Since the function $w_0(0,\cdot)$ is continuous on the compact set
$\T^3\setminus U_\delta(0)$ and has a unique minimum at $q=0$ there
exists $M>0$ such that $|w_0(0,q)|>M$ for all $q\in \T^3\setminus
U_{\delta}(0).$

Then by  $v(\cdot)\in \cB(\theta,\T^3)$ we have
\begin{equation}\label{D2}
|D_2(\zeta)-D_2(0)|\leq C \zeta^2,\,\zeta\in \R_+^0
\end{equation}
for some $C=C(\delta)>0.$

Applying the Morse lemma and computing some integrals  we obtain
that (see Lemma \ref{razlojeniya-1}) there exists a right-hand
derivative of $D_1(\cdot)$ at $\zeta=0$ and
\begin{align}\label{RhDer}
&\frac{\partial}{\partial \zeta }D_{1}(0)=
 \lim_{\zeta\to 0+}
  \frac{D_1(\zeta)-D_1(0)}{\zeta}=
 2\sqrt{2}\pi^2 l_1^{-\frac{3}{2}} v^2(0)
  (det W)^{-\frac{1}{2}}
\end{align}
and hence
\begin{equation}\label{D1-baho}
\big |D_1(\zeta)-D_1(0) \big |<C \zeta ,\quad \zeta\in \R^0_+
\end{equation}
holds for some positive $C.$

Then from \eqref{D2} and \eqref{D1-baho} it follows that the
right-hand derivative of $D(0,\cdot)$ at $\zeta=0$ exists and
$$
\frac{\partial}{\partial \zeta}D(0,0)=2\sqrt{2}\pi^2
l_1^{-\frac{3}{2}} v^2(0)
  (det W)^{-\frac{1}{2}}.
$$

Comparing \eqref{D2} and \eqref{D1-baho} we obtain \eqref{D-D}.

In the same way one can prove the inequality \eqref{partial D}.
\end{proof}

The following decomposition plays a important role in the proof of
the the main result, that is, the asymptotics  \eqref{asym.K}.

\begin{corollary}\label{razl.lemma.natijasi.} Let the operator $h(0)$ has an $m$
energy resonance. Then for any $p\in U_{\delta}(0),\delta>0$
sufficiently small, and
 $z\leq m$  the following decomposition
\begin{align*}
\Delta(p,z)= 2 \sqrt{2}\pi^2 v^2(0)l_1^{-\frac{3}{2}} (det W)^{-
\frac{1}{2}}\sqrt{m(p)-z}+\Delta^{(02)}(m(p)-z)+
\Delta^{(20)}(p,z)\nonumber
\end{align*}
holds, where $\Delta^{(02)}(m(p)-z)$ (resp. $\Delta^{(20)}(p,z)$) is
a function behaving like \\$O(({m(p)-z})^{\frac{1+\theta}{2}})$
(resp. $O(|p|^2)$) as $|{m(p)-z}| \to 0$ (resp. $p\to 0$ uniformly
in $z\leq m$).
\end{corollary}

\begin{proof} By Lemma \ref{h0 resonans} we have that
$\Delta(0,m)=0$ and $v(0)\neq 0$ and hence the proof of Corollary
\ref{razl.lemma.natijasi.} immediately follows from Lemma
\ref{razlojeniya} and equality \\$\Delta(p,z)=D(p,\sqrt{m(p)-z}).$
\end{proof}

\begin{lemma}\label{D.ineq} Let the operator $h(0)$ has an $m$
energy resonance. Then there exist positive numbers $c,C$ and
$\delta$ such that
\begin{equation}\label{c<(.,.)<c}
c |p| \leq \Delta(p,m) \leq C |p|\quad\mbox{for any}\quad p\in
U_\delta(0)
 \end{equation}
and
\begin{equation}\label{(.,.)>c}
\Delta(p,m) \geq c  \quad\mbox{for any}\quad\mbox p\in
\T^3\setminus U_\delta(0).
 \end{equation}
\end{lemma}
\begin{proof}
Corollary \ref{razl.lemma.natijasi.}  and the asymptotics (see part
$(ii)$ of Lemma \ref{minimum})
\begin{equation}\label{malfa}
m(p)=m+ \frac{l^2_1-l^2_2}{2l_1}(Wp,p)+o(p^3)\quad \mbox{as} \quad
p\to 0
\end{equation}
yields \eqref{c<(.,.)<c} for some positive numbers $c,C$.

The inequality \eqref{(.,.)>c} follows from the positivity (see
proof of Lemma \ref{inf}) and continuity of the function
$\Delta(\cdot,m)$ on the compact set $\T^3\setminus U_\delta(0).$
\end{proof}
 \begin{lemma}\label{main.ineq} Let the number $z=m$ is an
 eigenvalue of the operator $h(0).$
 Then there exist  numbers $\delta>0$ and $c>0$ such that
\begin{align*}
|\Delta(p,m)|\geq c p^2 \quad \mbox{for any}\quad p\in U_\delta(0)\\
|\Delta(p,m)|\geq c
 \quad \mbox{for all}\quad
p\in \T^3\setminus U_\delta(0).
\end{align*}
\end{lemma}
\begin{proof}
By Lemma \ref{zeroeigen} we have $\Delta(0,m)=0$ and $v(0)=0.$ Then
the function $\Delta(\cdot,m)$ can be represented in the form
$$
\Delta(p,m)=u(p)-u(0)+\frac{1}{2}(\Lambda(0,m)-\Lambda(p,m)).
$$
Then from Assumptions \ref{hypoth2} and \ref{hypoth3} it follows
that there exist positive numbers $\delta$ and $c$ such that the
statement of the lemma is fulfilled.
\end{proof}

\section{The essential spectrum of the operator $H$}

\subsection{The spectrum of the operator $\hat H$  }

We consider the operator $\hat{H}$ acting in\\
$\hat{\cH}=L_2({\T}^3)\oplus L_2(({\T}^3)^2)$ as
$$
\hat{H} \left( \begin{array}{ll}
f_1(p)\\
f_2(p,q)
\end{array} \right)=
\left( \begin{array}{ll}
u(p)f_1(p)+ \frac{1}{\sqrt{2}} \int\limits_{{\T}^3} v(q')f_2(p,q')dq'\\
\frac{1}{\sqrt{2}}v(q)f_1(p)+w_p(q)f_2(p,q)
\end{array} \right).
$$
The operator $\hat{H}$ commutes with any multiplication operator
$U_\gamma$ acting in  $\hat{\cH}$ as
$$
U_\gamma \left(
\begin{array}{ll}
f_1(p)\\
f_2(p,q)
\end{array} \right)=
\left( \begin{array}{ll}
\gamma(p)f_1(p)\\
\gamma(p)f_2(p,q)
\end{array} \right), \gamma \in L_2({\T}^3).
$$
Therefore the decomposition of the space $\hat{\cH}$ into the
direct integral
$$
\hat{\cH}= \int\limits_{{\T}^3} \oplus \cH^{(2)}dp
$$
yields the decomposition into the direct integral
\begin{equation}\label{decompose}
 \hat{H}= \int\limits_{{\T}^3} \oplus h(p)dp,
 \end{equation}
where the fiber operators $h(p),p\in\T^3$ are defined by (\ref{h}).

\begin{lemma}\label{spec}
For the spectrum  $\sigma(\hat H)$  of  $\hat H$ the equality
$$
\sigma(\hat H)\equiv\cup_{p\in {\T}^3} \sigma_d(h(p))\cup [m, M]
$$
holds.
\end{lemma}

\begin{proof} The assertion of the lemma follows from
the representation \eqref{decompose}
of the operator $\hat H$ and the theorem on decomposable operators
(see \cite{RSIV}).
\end{proof}
\begin{lemma}\label{ess.spec.H} The essential spectrum
$\sigma_{ess}(H)$ of the operator $H$ coincides with the spectrum of
$\hat H,$ that is,
\begin{equation}\label{essH}
 \sigma_{ess}(H)=\sigma(\hat H).
\end{equation}
\end{lemma}
\begin{proof}
In \cite{LRmn03} it has been proved that the essential spectrum
$\sigma_{ess}(H)$ of the operator $H$ coincides with $\sigma \cup
[m,M].$ By Lemma \ref{delta=0} we have
\begin{equation*}
 \sigma=\cup_{p \in {\T}^3}\sigma_d(h(p))
\end{equation*}
and hence by Lemma \ref{spec}  we obtain \eqref{essH}.
\end{proof}

\subsection{The structure of the essential spectrum of $H$}

Let us introduce the following notations:
\begin{align}\label{two.branch}
&\sigma_{two}(\hat H)=\cup_{p\in {\T}^3} \sigma_d(h(p)), \quad
a\equiv\inf \sigma_{two}(\hat H),\quad b\equiv\sup\sigma_{two}(\hat
H).
\end{align}

Since for any $p\in\T^3$ the function $\Delta(p,\cdot)$ is
decreasing in $(M,+\infty)$ by Lemma \ref{delta=0} the operator
$h(p),p\in \T^3$ either has  a unique eigenvalue in $(M,+\infty)$ or
not. It is easy to see that if $\Delta(p,M)\le 0$ for all $p\in
\T^3,$ then the operator $h(p),p\in \T^3$ has no eigenvalue in
$(M,+\infty).$

Thus the location and structure of the essential spectrum of $H$ can
be precisely described as well as in the following

\begin{theorem}\label{ess.loc} Let part $(i)$ of Assumption \ref{hypoth1}
 and Assumption
\ref{hypoth2} be
fulfilled and $\Delta(p,M)\le 0$ for all $p\in \T^3$.\\
(i) Assume that $\max\limits_{p\in\T^3}\Delta(p,m)<0.$ Then
\begin{equation*} \sigma_{ess}(H)= [a,b]\cup [m,
M]\quad \mbox{and}\quad b<m.
\end{equation*}
(ii) Assume that $\min\limits_{p\in\T^3}\Delta(p,m)<0$ and
$\max\limits_{p\in\T^3}\Delta(p,m)\ge 0.$ Then
\begin{equation*}
\sigma_{ess}(H)= [a, M]\quad \mbox{and}\quad a<m.
\end{equation*}
(iii) Assume that $\min\limits_{p\in\T^3}\Delta(p,m)\ge 0.$ Then
$$
\sigma_{ess}(H)= [m,M]. $$
\end{theorem}

\begin{proof}
$(i)$ Let $\max\limits_{p\in\T^3}\Delta(p,m)<0$. Then by Lemma
\ref{neg eigen} for all $p\in \T^3$ the operator $h(p)$ has a unique
eigenvalue $z(p)<m.$

By part $(i)$ of Assumption \ref{hypoth1} and Assumption
\ref{hypoth2} and Lemma \ref{delta=0} the map $z: p\in  \T^3 \to
z(p)$ is of class $C^{(2)}(\T^3).$

Therefore the range $\Im z$ of the function $z(\cdot)$ is a
connected closed subset of $(-\infty,m)$, that is, $\Im z=[a,b]$ and
$b<m$ and hence $\sigma_{two}(\hat H)=[a,b].$

$(ii)$ Let $\min\limits_{p\in\T^3}\Delta(p,m)<0$ and
$\max\limits_{p\in\T^3}\Delta(p,m)\ge 0$. Then by assertion $(ii)$
of Lemma \ref{neg eigen} there exists a non void  open set $D$ such
that for any $p\in D $ the operator $h(p)$ has a unique eigenvalue
$z(p)<m.$

Since for any $p\in \T^3$ the operator $h(p)$ is bounded and $\T^3$
is compact set, there exists a positive number $C$ such that
$\sup\limits_{p\in \T^3}||h(p)||\le C$ and for any $p\in \T^3$ we
have
\begin{equation}\label{spectrsubset}
\sigma(h(p))\subset [-C,C].
\end{equation}

For any $q\in \partial D=\{p\in \T^3: \Delta(p,m)=0\}$ there exist
$\{p_n\}\subset D$ such that $p_n\to q$ as $n\to \infty.$ Set
$z_n=z(p_n).$ Then by Lemma \ref{neg eigen} for any $p_n\in D$ the
inequality $z_n<m$ holds and from \eqref{spectrsubset} we get
$\{z_n\}\subset [-C,m].$ Without loss of generality (otherwise we
would have to take a subsequence) we assume that $z_{n}\to z_0$ as
$n\to \infty$ for some $z_0\in [-C,m].$

From the continuity of the function $\Delta(\cdot,\cdot)$ in
$\T^3\times (-\infty,m]$ and $p_n\to q$ and $z_{n}\to z_0$ as $n\to
\infty$ it follows that
$$
0=\lim\limits_{n\to \infty}\Delta(p_{n},z_{n}) =\Delta(q,z_0).
$$

Since for any $p\in \T^3$ the function $\Delta(p,\cdot)$ is
decreasing  in $(-\infty,m]$ and $q\in
\partial D$ we see that $\Delta(q,z_0)=0$ if and only if
$z_0=m.$

For any $q\in \partial D$ we define
$$
z(q)=\lim\limits_{p\to q, p\in D}z(p) =m.
$$

Since the function $z(\cdot)$ is continuous on the compact set
$D\cup\partial D$ and $z(q)=m$ for all $q\in
\partial D$ we conclude that
$$
\Im z=[a,m],\quad a<m.
$$

Hence the set
$$
\{z \in \sigma_{two}(\hat H): z\le m\}= \cup_{p\in \T^3}
\sigma_d(h(p))\cap (-\infty,m]
$$
coincides with the set $\Im z=[a,m].$ Then Lemma \ref{spec}
completes the proof of $(ii).$

$(iii)$ Let $\min\limits_{p\in\T^3}\Delta(p,m)\ge 0$. Then by Lemma
\ref{neg eigen} for all $p\in \T^3$ the operator $h(p)$ has no
eigenvalues lying on the l.h.s. of $m.$

Hence we have
\begin{equation*}
 \sigma(\hat H)=[m,M].
\end{equation*}

So Lemma \ref{ess.spec.H} complete the proof of Theorem
\ref{ess.loc}
\end{proof}

\section{The Birman-Schwinger principle}
In this section we prove an analogue of the Birman-Schwinger
principle.

For a bounded self-adjoint operator $A,$ we define $n(\lambda,A)$ as
$$
n(\lambda,A)=sup\{ dim F: (Au,u) > \lambda,\, u\in F,\,||u||=1\}.
$$
The number $n(\lambda,A)$ is equal the infinity if $\lambda$ is in
the essential spectrum of $A$ and if $n(\lambda,A)$ is finite, it is
equal to the number of the eigenvalues of $A$ bigger than $\lambda$.
By the definition of $N(z)$ we have
$$ N(z)=n(-z,-H),\,-z
> -\tau_{ess}(H).
$$

Since the function $\Delta(\cdot,\cdot)$ is positive on $ \T^3\times
(-\infty,\tau_{ess}(H))$ there exists a positive square root of
$\Delta(p,z)$ for any  $p\in\T^3$ and $z<\tau_{ess}(H).$

In our analysis of the spectrum of $H$ the crucial role is  played
by the compact operator  $ T(z), z<\tau_{ess}(H)$ in the space
${\cH}^{(2)}$ with the entries
$$
(T_{00}(z)f_0)_0=(1-u_0-z)f_0,\quad
(T_{01}(z)f_1)_0=-\int\limits_{{\T}^3} \frac{b(q')f(q')
dq'}{\sqrt{\Delta(q',z)}},
$$
$$
T_{10}(z)=T_{01}^*(z),\quad
(T_{11}(z)f_1)_1(p)=\frac{b(p)}{2\sqrt{\Delta(p,z)}}\int\limits_{{\T}^3}
\frac{b(q')f(q') dq'}{\sqrt{\Delta(q',z)}(w(p,q')-z)}.
$$

The following lemma is a realization of the well known
Birman-Schwinger principle for the operator $H$ (see
\cite{ALzMahp04,Sob,Tam94}).
\begin{lemma}\label{b-s} For $z<\tau_{ess}(H)$
the operator $T(z)$ is compact and continuous in $z$ and
\begin{equation}\label{N=n}
N(z)=n(1,T(z)).
\end{equation}
\end{lemma}

\begin{proof}
The operator $H$ can be decomposed as
$$
H=\left(
\begin{array}{ccc}
H_{00} & 0 & 0\\
0 & H_{11} & 0\\
0 & 0 & H_{22}\\
\end{array}
\right)+  \left(
\begin{array}{ccc}
0 & H_{01} & 0\\
H_{10} & 0 & H_{12}\\
0 & H_{21} & 0\\
\end{array}
\right).
$$

The operator $H_{ii}-z{\bf I},i=1,2$ is positive and invertible for
$z<\tau_{ess}(H),$ where ${\bf I}$ is the identity operator on
$\cH.$ Hence there exist a square root of $R_{i}(z)=(H_{ii}-z{\bf
I})^{-1},\,i=1,2.$ Then one has $f \in \cH$ and $((H-z {\bf
I})f,f)<0$ if and only if $((M(z)-{\bf I})g,g)>0$ and
$g_0=f_0,\,g_i=(H_{ii}-z)^{\frac{1}{2}}f_i,i=1,2,$
 the operator
$M(z),z<\tau_{ess}(H)$ has the entries
$$
(M_{00}(z)f_0)_0=(1-u_0+z)f_0,\,\, M_{01}(z)=-H_{01}
R_1^{\frac{1}{2}}(z),\,\, M_{12}(z)=-R_1^{\frac{1}{2}}(z) H_{12}
R_2^{\frac{1}{2}}(z),
$$
$$
M_{10}(z)=M_{01}^*(z),\quad M_{21}(z)=M_{12}^*(z),
$$
otherwise
$$
M_{\alpha\beta}(z)=0.
$$

It follows that
\begin{equation}\label{N(z)=}
N(z)=n(1,M(z)).
\end{equation}

Let $V(z),z<\tau_{ess}(H)$ be the operator in $\cH^{(2)}$ with the
entries
$$
V_{11}(z)=R_1^{\frac{1}{2}}(z) H_{12} R_2(z) H_{21}
R_1^{\frac{1}{2}}(z),\quad \mbox{otherwise} \quad
V_{\alpha\beta}(z)=M_{\alpha\beta}(z),\,\alpha,\beta=0,1.
$$

A direct calculation shows that
\begin{equation}\label{n(1,M(z))=}
n(1,M(z))=n(1,V(z)).
\end{equation}

One has $\cH^{(2)}$ and $((V(z)-I)\varphi,\varphi)>0$ iff
$\psi_0=\varphi_0$ and $\psi_1=R_1^{\frac{1}{2}}(z)\varphi_1$ and
\begin{equation*}
(\psi_0,\psi_0)_0+((H_{11}-z)\psi_1,\psi_1)_1<
(M_{00}(z)\psi_0,\psi_0)_0-
\end{equation*}
\begin{equation}\label{n(1,G(z))}
-(H_{01}\psi_1,\psi_0)_0-(H_{10}(z)\psi_0,\psi_1)_1+
(H_{12}R_2(z)H_{21}\psi_1,\psi_1)_1,
\end{equation}
where ${I}$ is the identity operator on $\cH^{(2)}.$ This means that
\begin{equation}\label{n(1,V(z))=}
n(1,V(z))=n(-z, G(z)),
\end{equation}
where
\begin{equation*}
G(z)=\left( \begin{array}{cc}
-H_{00}-2z & -H_{01}\\
-H_{10} & H_{12}R_2(z)H_{21}-H_{11}\\
\end{array} \right),
\end{equation*}

Now we represent the operator $H_{21}$ as a sum of two operators
$H^{(1)}_{21}$ and $H^{(2)}_{21}$ acting from $L_2(\T^3)$ to
$L_2((\T^3)^2)$ as
$$
(H_{21}^{(1)}f_1)(p,q)=\frac{1}{2} v(p)
f_1(q),\,(H_{21}^{(2)}f_1)(p,q)= \frac{1}{2} v(q)f_1(p).
$$

Since $z<\tau_{ess}(H)$   for any $p\in \T^3$ the function
${\Delta}( p, \cdot)$ is positive and hence the operator
$H_{11}-z-H_{12}R_2(z)H^{(2)}_{21}$ is positive and invertible and
$$
(H_{11}-z-H_{12}R_2(z)H^{(2)}_{21})^{-\frac{1}{2}}=R_{11}^{\frac{1}{2}}(z).
$$

Now $\psi\in \cH^{(2)}$ and the inequality \eqref{n(1,G(z))} holds
iff $\eta_0=\psi_0,\,\eta_1=R_1^{\frac{1}{2}}(z)\varphi_1$ and
$((T(z)-I)\eta,\eta)>0$ and hence
\begin{equation}\label{n(1,T(z))=}
n(-z,G(z))=n(1,T(z)).
\end{equation}
The equalities \eqref{N(z)=}, \,\eqref{n(1,M(z))=},\,
\eqref{n(1,V(z))=} and \eqref{n(1,T(z))=} give \eqref{N=n}.

Finally we note that for any  $z<\tau_{ess}(H)$ the operator $T(z)$
is compact and continuous in $z.$
\end{proof}

\section{ The finiteness of the  number of eigenvalues of the  operator $H$.}

\begin {lemma}\label{G-S} Let the conditions in part $(i)$ of Theorem \ref{fin} be fulfilled.
 Then for any $z\leq m$  the operator $T(z)$ is compact and
is continuous from the left up to $z=m.$
\end{lemma}
\begin{proof}
Denote by $Q(p,q;z)$ the kernel of the operator $T_{11}(z),\,z\leq
m,$ that is,
$$
Q(p,q;z)=\frac{b(p)b(q))}{2\sqrt{\Delta(p,z)}(w(p,q)-z)\sqrt{\Delta(q,z)}}.
$$
Since the function $v\in \cB (\theta,\T^3)$ is even and $v(0)=0$ we
have $|v(p)|\leq C|p|^\theta$ for any $p\in\T^3$ and for some $C>0$.
By virtue of Lemmas \ref{main.ineq}  and \ref{U.ineq} the kernel
$Q(p,q;z)$ is estimated by
$$ C
 (\frac{\chi_{\delta}(p)}{|p|}+1)(
\frac{|p|^\theta|q|^\theta
\chi_{\delta}(p)\chi_{\delta}(q)}{p^2+q^2}+1
 )(\frac{\chi_{\delta}(q)}{|q|}+1)
 ,
$$ where  $ \chi_{\delta}(p)$ is the characteristic function of
$U_\delta(0).$

The latter function is square-integrable on $(\T^3)^2$ and hence we
have that, for any $ z\leq m,$  $T_{11}(z)$ is a Hilbert-Schmidt
operator.

The kernel function of $T_{11}(z)$ is continuous in $p,q \in
\T^3,\,z<m$ and square-integrable
 on $(\T^3)^2$ for $z\leq m.$
Now the continuity of the operator $T_{11}(z)$ from the left up to
$z=m$ follows from Lebesgue's dominated convergence theorem.

Since for all $z\leq m$ the operators $T_{00}(z), $ $T_{01}(z)$ and
$T_{10}(z)$ are of rank $1$ and continuous from the left up to $z=m$
we can conclude that $T(z)$ is compact and continuous from the left
up to $z=m.$
\end{proof}

We are now ready for the

 {\bf Proof of $(i)$ of Theorem \ref{fin}.} Let the conditions
 in part $(i)$ of Theorem \ref{fin} be fulfilled.
By Lemma \ref{b-s}  we have
$$ N(z)=n(1,T(z)),\,\mbox{as}\,\,z<m $$ and by Lemma \ref{G-S}
for any $\gamma\in [0,1)$ the number $n(1-\gamma,T(m)) $ is finite.
Then we have
$$
n(1,T(z))\leq n(1-\gamma,T(m))+n(\gamma,T(z)-T(m))
$$
for all $z<m$ and $\gamma \in (0,1).$ This relation can  easily be
obtained by a use of the Weyl inequality
$$
n(\lambda_1+\lambda_2,A_1+A_2)\leq n(\lambda_1,A_1)+n(\lambda_2,A_2)
$$
for the sum of compact operators $A_1$ and $A_2$ and for any
positive numbers $\lambda_1$ and $\lambda_2.$

Since $T(z)$ is continuous from the left up to $z=m$, we obtain $$
\lim_{z\to m-0} N(z)= N(m)\leq n(1-\gamma,T(m))\,\, \mbox{for
all}\,\, \gamma \in (0,1). $$

Thus
\begin{equation}\label{N(0)finite}
 N(m)\leq n(1-\gamma,T(m))<\infty.
\end{equation}
 The inequality \eqref{N(0)finite} proves the assertion $(i)$ of Theorem \ref{fin}.
\qed

\section{Asymptotics for the number of
eigenvalues of the  operator $H$.}

In this section we shall closely follow the A. Sobolev method
\cite{Sob} to derive the asymptotics \eqref{asym.K} for the number
of eigenvalues of $H$.

We shall first establish the asymptotics of $n(1,T(z))$ as $z\to m
-0.$ Then part $(ii)$ Theorem \ref{fin} will be deduced by a
perturbation argument based on the following lemma.

 \begin{lemma}\label{comp.pert}
 Let $A (z)=A_0 (z)+A_1 (z),$ where $A_0(z)$ $(A_1(z))$ is
compact and continuous in $z<m.$  Assume that for some function
$f(\cdot),\,\, f(z)\to 0,\,\, z\to -0$ there exists the
$$
\lim_{z\rightarrow -0}f(z)n(\lambda,A_0 (z))=l(\lambda),
$$
and is continuous in $\lambda>0.$ Then the same limit exists for
$A(z)$ and
$$ \lim_{z\rightarrow -0}f(z)n(\lambda,A (z))=l(\lambda).
$$
\end{lemma}
For the proof of Lemma \ref{comp.pert}, see Lemma 4.9 of \cite{Sob}.

\begin{remark} Since $\cU(\cdot)$ is continuous in $\mu>0,$ according
to Lemma \ref{comp.pert} any perturbations of the operator $A_0(z)$
defined in Lemma \ref{comp.pert}, which is compact and continuous up
to $z=m$ do not contribute to the asymptotics \eqref{asym.K}. During
the Section 7 we use this fact without further comments.
\end{remark}

By Assumption \ref{hypoth1} we get
\begin{equation}\label{asymp1}
w(p,q)=m+\frac{1}{2}\big(
l_1(Wp,p)+2l_2(Wp,q)+l_1(Wq,q))+O(|p|^{3+\theta}+|q|^{3+\theta})\,\,
\text{as}\,\, p,q\rightarrow 0.
\end{equation}
By the representation \eqref{malfa} and Corollary
\ref{razl.lemma.natijasi.} we get
\begin{equation}\label{asymp2}
\Delta(p,z) = \frac{4\pi^2 v^2(0)}
 {l_1^{{3}/{2}} (\det W)^{\frac{1}{2}}}
 \left [ l (Wp,p) -2(z-m) \right ]^{\frac{1}{2}}+
 O((|p|^2+|z-m|)^{\frac{1+\theta}{2}})
 \end{equation}
as $p,|z-m|\rightarrow 0, $ where $ l={(l^2_1-l^2_2)}/{l_1}. $

Denote by $\hat \chi_\delta(\cdot)$ the  characteristic function of
$\hat U_\delta(0)=\{ p\in \T^3:\,\, |W^{\frac{1}{2}}p|<\delta \}.$

 Let $T(\delta;|z-m|)$ be the
operator in $\cH^{(2)}$ defined by

\begin{equation*}
T(\delta;|z-m|)=\left( \begin{array}{cc}
0 & 0\\
0 & T_{11}(\delta;|z-m|)\\
\end{array} \right),
\end{equation*}
where the $ T_{11}(\delta;|z-m|)$ is the integral operator in
$\cH_1$ with the kernel
\begin{align*}
\frac{l_1^{\frac{3}{2}}({\det W})^{\frac{1}{2}}\hat \chi_\delta
(p)\hat \chi_\delta (q)(l(Wq,q)+ 2|z-m|)^ {-1/4}} {2\pi^2(l(Wp,p)+
2|z-m|)^{1/4}(l_1(Wp,p)+ 2l_2(Wp,q)+l_1(Wq,q)+2|z-m|)}.
\end{align*}

The main technical point to apply Lemma \ref{comp.pert} is the
following
\begin{lemma}\label{H-Sh} Let the conditions
in part $(ii)$ of Theorem \ref{fin} be fulfilled. Then the operator
$ T(z)-T(\delta; |z-m|)$ is compact and is continuous in $z\leq m.$
\end{lemma}
\begin{proof}
Applying the asymptotics \eqref{asymp1}, \eqref{asymp2} and Lemmas
\ref{D.ineq} and \ref{U.ineq} one can estimate the kernel of the
operator $ T_{11}(z)-T_{11}(\delta; |z-m|)$ by
\begin{equation*}
 C [ \frac{|p|^\theta}{|p|^{\frac{1}{2}}} +\frac{|q|^\theta}{|q|^{\frac{1}{2}}}+
\frac{|p|^{\theta}+|q|^{\theta}}{|p|^{\frac{1}{2}}(p^2+q^2)|q|^{\frac{1}{2}}}+
\frac{|m-z|^{\frac{\theta}{2}}(p^2+q^2)^{-1}}{(|p|^2+|m-z|)^{\frac{1}{4}}
(|q|^2+|m-z|)^{\frac{1}{4}}}+1 ]
\end{equation*}
 and hence
the operator $ T_{11}(z)-T_{11}(\delta; |z-m|)$
 belongs to the Hilbert-Schmidt class for all
$z \leq m.$ In combination with the continuity of the kernel of the
operator in  $z<m$ this  gives   the continuity of $
T_{11}(z)-T_{11}(\delta;|z-m|)$ in  $z\leq m.$

It is easy to see that $T_{00}(z), $ $T_{01}(z)$ and $T_{10}(z)$ are
 rank 1 operators and they are continuous from the left up to $z=m.$
The details are omitted.
\end{proof}

Let us now recall some results from \cite{Sob}, which are important
in our work. \\ Let ${\bf S}_{{\bf r}}:L_2((0,{\bf r}),
{\sigma_0})\to L_2((0,{\bf r}),{\sigma_0}) $ be the integral
operator with the kernel
\begin{equation}\label{S kernel}
 S(y;t)=(2\pi)^{-2}\frac{l_0}
{\cosh y+st}
\end{equation}
and
\begin{align*}
{\bf r}=1/2 | \log |z-m||,\,y=x-x',\,t=<\xi, \eta>,\,\xi, \eta \in
\S^2,\,l_0=\big( \frac{l^2_1}{l^2_1-l^2_2} \big)^{\frac{1}{2}},\,
 s=\frac{l_2} {l_1},
\end{align*}
${\sigma_0}=L_2(\S^2),\, \S^2$ being the unit sphere in $\R^3.$

The coefficient in the asymptotics of $N(z)$ will be expressed by
means of the self-adjoint integral operator $\hat{\bf
S}(\lambda),\,\lambda\in \R,$ in the space $L_2(\S^2)$ whose kernel
depends on the scalar product $t=<\xi,\eta>$ of the arguments
$\xi,\eta\in\S^2$ and has the form
\begin{equation*}
\hat {\bf S}(t;\lambda)=(2\pi)^{-1}l_0\frac{\sinh[\lambda(arccos
st)]} {(1-s^2t^2)^{\frac{1}{2}}\sinh (\pi\lambda)}.
\end{equation*}

For $\mu>0,$ define
$$
{U}(\mu)= (4\pi)^{-1} \int\limits_{-\infty}^{+\infty} n(\mu,\hat{\bf
S}(y))dy.
$$

Set $\cU_0=U(1).$

The following lemma can be proved in the same way as Theorem 4.5
in \cite{Sob}.
\begin{lemma}\label{lim} Let ${\bf
S}_{\bf r}$ be the operator defined in \eqref{S kernel}. Then for
any $\mu>0$ the equality
$$ \lim\limits_{{\bf r}\to \infty} \frac{1}{2}{\bf
r}^{-1}n(\mu,{\bf S}_{\bf r})={U}(\mu) $$ holds.
\end{lemma}

The following theorem is basic for the proof of the asymptotics
\eqref{asym.K}.
\begin{theorem}\label{main} Let  the conditions in part $(ii)$ of Theorem \ref{fin}
be fulfilled. Then the equality
$$ \lim\limits_{|z-m|\to 0} \frac{n(1,T(z))} {|log|z-m||}
=\lim\limits_{{\bf r}\to \infty} \frac{1}{2}{\bf r}^{-1}n(1,{\bf
S}_{\bf r}) $$ holds.
\end{theorem}

\begin{proof} As in Lemma \ref{H-Sh}, it can be shown that $
T(z)-T(\delta; |z-m|)$ defines a compact operator continuous in
$z\le m$  and this does not contribute to the asymptotics
\eqref{asym.K}.

The space of functions  having support in $\hat U_\delta(0)$  is an
invariant subspace for the operator $T_{11}(\delta;|z-m|).$

Let  $\hat T_{11}^{(0)}(\delta;|z-m|)$ be the restriction of the
operator $T_{11}(\delta;|z-m|)$ to the subspace $L_2(\hat
U_{\delta}(0)).$ One verifies  that the operator $\hat
T_{11}^{(0)}(\delta;|z-m|)$ is unitary equivalent to the following
operator $T_{11}^{(0)}(\delta;|z-m|)$ acting in $L_2(\hat
U_{\delta}(0))$ as
\begin{align*}
(T_{11}^{(0)}(\delta;|z-m|)f)(p)= \frac{l_1^{{3}/{2}}}{2 \pi^2}
\int\limits_{U_\delta(0)} \frac{ (l p^2+ 2|z-m|)^{-1/4} (l q^2+
2|z-m|)^ {-1/4} } {l_1 p^2+ 2l_2(p,q)+l_1 q^2 +2|z-m|} f(q)d q.
\end{align*}

Here the equivalence is performed by the unitary dilation $${\bf
Y}:L_2( U_{\delta}(0))
 \to L_2(\hat U_{\delta}(0)),\quad
 ({\bf Y} f)(p)=f(U^{-\frac{1}{2}}p).
 $$

The operator $T_{11}^{(0)}(\delta;|z-m|)$ is unitary equivalent
 to the integral operator $T_{11}^{(1)}(\delta;|z-m|):L_2(U_r(0)) \to L_2(U_r(0))$
 with the kernel
\begin{align*}
\frac{l_1^{{3}/{2}}}{2 \pi^2} \frac{  (l p^2+ 2)^{-1/4} (l q^2+ 2)^
{-1/4} } {l_1 p^2+ 2l_2(p,q)+l_1 q^2 +2},
\end{align*}
where $r=|z-m|^{-\frac{1}{2}}$ and  $U_r(0)=\{ p\in \R^3:|p|<r\}.$

The equivalence of the operators $T_{11}^{(0)}(\delta;|z-m|)$ and
$T_{11}^{(1)}(\delta;|z-m|)$ is performed by the unitary dilation
 $${\bf B}_r:L_2
(U_\delta(0)) \to L_2(U_r(0)),\quad
 ({\bf B}_r f)(p)=(\frac{r}{\delta})^{-3/2}f(\frac{\delta}
{r}p).$$ Further, we may replace $$(l p^2+2)^{-1/4},\, (l
q^2+2)^{-1/4} \quad \mbox{ and}\quad l_1 p^2+ 2l_2(p,q)+l_1 q^2
 +2$$
 by
$$(l p^2)^{-1/4}(1-\chi_1(p)),\,\, (l
q^2)^{-1/4}(1-\chi_1(q))
 \quad \mbox{ and}\quad
l_1 p^2+ 2l_2(p,q)+l_1 q^2 ,$$
  respectively, since the error will be
a Hilbert-Schmidt operator  continuous up to  $z=m,$ where
$\chi_1(\cdot)$ is a characteristic function of the ball $U_1(0).$
Then we get the integral operator $T_{11}^{(2)}(r)$ on $L_2(U_r(0)
\setminus U_1(0))$  with the kernel
\begin{align*}
l^{-\frac{1}{2}} \frac{l_1^{{3}/{2}}}{2 \pi^2} \frac{|p|^{-1/2}|
q|^{-1/2}} {l_1  p^2+ 2l_2(p,q)+l_1 q^2}.
\end{align*}

By the dilation $${\bf M}:L_2(U_r(0) \setminus U_1(0))
\longrightarrow L_2((0,{\bf r})\times {\sigma_0}),$$
 where
$(M\,f)(x,w)=e^{3x/2}f(e^{ x}w),\, x\in (0,{\bf r}),\, w \in
{\S}^2,$ one sees that the operator $T_{11}^{(2)}(r)$ is unitary
equivalent to the integral operator ${\bf S}_{{\bf r}}.$
\end{proof}

{\bf Proof of $(ii)$ of Theorem \ref{fin}.} Let the conditions in
 part $(ii)$ of Theorem \ref{fin} be fulfilled.

Similarly to \cite{Sob} we can show that
\begin{equation}\label{Slambda}
\cU_0={U}(1) \ge \frac{1}{4\pi} \int\limits_{-\infty}^{+\infty}
n(1,\hat{\bf S}^{(0)}(y))dy\ge\frac{1}{4\pi} mes\{y:\hat{S}^{(0)}(y)>1\},
\end{equation}
where $\hat{\bf S}^{(0)}(y)$  is the multiplication operator by the
number
\begin{equation*}
\hat {S}^{(0)}(y)=l_0\frac{\sinh(y arcsin s)} {sy\cosh \frac{\pi
y}{2}}
\end{equation*}
in the subspace of the harmonics of degree zero.

The positivity of $\cU_0$ follows  from the facts that $l_0>1,\,\hat
{S}^{(0)}(0)>1$ and continuity of $\hat {S}^{(0)}(y).$ Taking into
account the inequality \eqref{Slambda} and Lemmas \ref{b-s},
 \ref{lim},\,\ref{main},  we complete the proof of $(ii)$ of Theorem
\ref{fin}. \qed

\appendix
\section{}
\begin{lemma}\label{examp}
Let the function $v$ as in Assumption \ref{hypoth2} and the function
$w$ be defined by \eqref{u va w} and $\varepsilon$ be a
real-analytic conditionally negative definite function on ${\T}^3$
with a unique non-degenerate minimum at the origin. Then Assumption
\ref{hypoth3} is fulfilled
\end{lemma}
\begin{proof}
It is known that the conditionally negative definite function
$\varepsilon$ admits the (L\'evy-Khinchin) representation (see,
e.g., \cite{BCR})
$$
\varepsilon(p)=\varepsilon(0)+\sum_{s\in
\Z^d\setminus\{0\}}(e^{\mathrm{i}(p,s)}-1)\hat \varepsilon (s),\quad
p\in \T^3,
$$
which is equivalent to the requirement that  the Fourier
coefficients $\hat \varepsilon(s)$ with $s\ne 0$   are non-positive,
that is,
\begin{equation}\label{non-neg}
 \hat \varepsilon(s)\le 0, \quad s\ne 0,
\end{equation}
and the series $\sum_{s\in \Z^3\setminus\{0\}}\hat \varepsilon (s)$
converges absolutely.

If the function $\varepsilon(\cdot)$ is real-valued, then the
equality
\begin{equation}\label{even}
\hat \varepsilon (s)=\hat \varepsilon (-s),\quad s\in \Z^3
\end{equation}
holds and hence
\begin{equation}\label{Lev.Khi1}
 \varepsilon(p)=\varepsilon(0)+\sum_{s\in
\Z^3\setminus\{0\}}\hat \varepsilon (s)(\cos(p,s)-1),\quad p\in
\T^3.
\end{equation}

Thus $\varepsilon(\cdot)$ is even on $\T^3.$

Since $w$ and $v$ are even the function $\Lambda(\cdot)$ is also
even. Then from the equality
\begin{equation*}
w_0(t)-\frac{w_p(t)+w_p(-t)}{2}= \sum_{s\in
Z^3\setminus\{0\}}\hat{\varepsilon}(s)(1+\cos (q,s))(1-\cos (p,s))
\end{equation*}
we get
\begin{equation}\label{Lamb-lamb}
\Lambda(0,m)-\Lambda(p,m)=\sum_{s\in
Z^3\setminus\{0\}}(-\hat{\varepsilon}(s))(1-\cos
(p,s))B(p,s)+\tilde{B}(p),
\end{equation}
where
\begin{equation*}
B(p,s)=\frac{1}{2}\int\limits_{{\T}^3}
\frac{(1+cos(q,s))[w_p(t)+w_{-p}(t)-2m]}
{(w_p(t)-m)(w_{-p}(t)-m)(w_0(t)-m)}v^2(t)dt,
\end{equation*}
$$
\tilde{B}(p)=\frac{1}{4}\int\limits_{{\T}^3} \frac{
[w_p(t)-w_{-p}(t)]^2} {(w_p(t)-m)(w_{-p}(t)-m)(w_0(t)-m)}v^2(t)dt.
$$

Let $\delta>0$ such that $mes\{ (\T^3\setminus U_\delta(0))\cap
supp\,\, v\}>0$ holds. We rewrite the function $B(p,s)$ as a sum of
two functions
\begin{equation*}
B_1(p,s)=\frac{1}{2}\int\limits_{{\T}^3\setminus U_\delta(0)}
\frac{(1+cos(q,s))[w_p(t)+w_{-p}(t)-2m]}
{(w_p(t)-m)(w_{-p}(t)-m)(w_0(t)-m)}v^2(t)dt
\end{equation*}
and
\begin{equation*}
B_2(p,s)=\frac{1}{2}\int\limits_{U_\delta(0)}
\frac{(1+cos(q,s))[w_p(t)+w_{-p}(t)-2m]}
{(w_p(t)-m)(w_{-p}(t)-m)(w_0(t)-m)}v^2(t)dt.
\end{equation*}

Since the function $w$ has a unique minimum at $(0,0)$ and $v\in \cB
(\theta,\T^3)$ we obtain that for all $p\in \T^3$ the function
$$
F(p,\cdot)=\frac{(1-\chi_\delta(\cdot))[w_p(\cdot)+w_{-p}(\cdot)-2m]}
{(w_p(\cdot)-m)(w_{-p}(\cdot)-m)(w_0(\cdot)-m)}v^2(\cdot)
$$
belongs to $L_1(\T^3),$ where $\chi_\delta(\cdot)$ is the
characteristic function of $ U_\delta(0).$

Then by the Riman-Lebesque theorem for all $p\in \T^3$ we have
$$
\int\limits_{{\T}^3} cos(t,s)F(p,t)dt\to 0\quad \mbox{as}\quad s\to
\infty
$$
and hence the inequalities
$$ mes \{ (\T^3\setminus U_\delta(0))\cap supp\,\,v(\cdot)\}>0$$ and
$$F(p,t)>0\quad\mbox{for\, all}\quad p\in \T^3\quad \mbox{and \,a.e.} \quad
t\in (\T^3\setminus U_\delta(0))\cap supp\,\,v(\cdot)$$ implies that
$$
B_1(p,s)\to \int\limits_{{\T}^3} F(p,t)dt>0\quad \mbox{as}\quad s\to
\infty.
$$

Then from continuity of the function
$$
\int\limits_{{\T}^3} F(p,t)dt
$$
on the compact set $\T^3$ it follows that there exists $C_1>0$
such that $B_1(p,s)>C_1,\,s\in \Z^3,\,p\in \T^3.$ Since
$B_2(p,s)\geq 0,\,s\in \Z^3,\,p\in \T^3$ we have $B(p,s)>C_1,\,s\in
\Z^3,\,p\in \T^3.$ Then according to    $\tilde{B}(p)\geq 0,\, p\in
\T^3$  and $\hat{\varepsilon}(s)\leq 0,\,s\in \Z^3\setminus\{0\}$
(see \eqref{non-neg}) the representations \eqref{Lev.Khi1} and
\eqref{Lamb-lamb} implies that
\begin{equation*}
\Lambda(0,m)-\Lambda(p,m)\geq C_1 (\varepsilon(p)-\varepsilon(0))
\end{equation*}
and since the function $\varepsilon$ has a unique non-degenerate
minimum at the origin we obtain that
\begin{equation}\label{Lambda}
\Lambda(0,m)-\Lambda(p,m)\geq C_2p^2
\end{equation}
for some $C_2>0.$ The inequality \eqref{Lambda}  completes the proof
of the lemma.
\end{proof}
\begin{lemma}\label{minimum}
Let Assumption \ref{hypoth1} be fulfilled. Then there exists a $ {
\delta } $-neighborhood $U_ {\delta } (0)\subset \T^3$ of the point
$p=0$ and a function $q_0 (\cdot)\in C^{(2)}(U_ {\delta
} (0))$   such that:\\
(i)   for any $p { \in } U_ { \delta }(0)$  the point
  $q_0(p) $ is a unique non-degenerate minimum  of $w_p(\cdot)$ and
\begin{equation}\label{q0 asym}
q_0(p)=-\frac{l_2}{l_1}p+O(|p|^{2+\theta})\,\,as\,\,p \to 0.
\end{equation}\\
(ii) The function $m(\cdot)= w(\cdot,q_0(\cdot))$ is of class $q_0
(\cdot)\in C^{(3)}(U_ {\delta } (0))$ and has the asymptotics
 \begin{equation}\label{min.raz}
m(p)=m+\frac{l^2_1-l^2_2}{2l_1}(Wp,p)+O(|p|^{3+\theta}) \quad
\mbox{as}\quad p \to 0.
\end{equation}
\end{lemma}
\begin{proof}
$(i)$ The existence the number ${\delta }>0$ and a function $q_0
(\cdot)\in C^{(2)}(U_ {\delta } (0))$  such that for all $p { \in }
U_ { \delta} ( 0 ) $ the point $ q_0(p)$ is the unique
non-degenerate minimum point of $w_p(\cdot)$ can be proven by the
same way as Lemma 3 in \cite{Ltmf92}.

The    function $q_0(\cdot)$ is an odd function in $p \in
U_\delta(0).$

Indeed, since $w(\cdot,\cdot)$ is even with respect $(p,q)$ for all
$p \in U_\delta(0)$ we obtain
\begin{equation}\label{q0}
w_{-p}(-q_0(p))= m(p)=m(-p)= w_{-p}(q_0(-p)).
\end{equation}

Since for each $p \in U_\delta(0)$ the point $q_0(p)$ is the unique
non-degenerate  minimum of the function $w_{p}(\cdot)$ by \eqref{q0}
we have $ q_0(-p)=-q_0(p),\,p\in U_\delta(0).$

The asymptotics \eqref{q0 asym} follows from the fact that
$q_0(\cdot)$ is an odd function.  The coefficient $-\frac{l_2}{l_1}$
is calculated using identity ${\bigtriangledown } w(p, q_0(p))\equiv
0,\,p\in U_\delta(0).$

$(ii)$ By asymptotics \eqref{asymp1}, \eqref{q0 asym} and the
equality $m(p)=w_{p}(q_0(p))$  we obtain asymptotics
\eqref{min.raz}.
\end{proof}

\begin{lemma}\label{inf}
If the conditions of Theorem \ref{fin} are fulfilled, then  the
operator $h(p),\,p\in \T^3$ has no eigenvalues lying on l.h.s. of $
$ $m.$ Therefore, $\inf \sigma_{ess}(H)=\inf
\sigma_{three}(H)=\inf\sigma_{two}(H)= m.$
\end{lemma}
\begin{proof} It suffices to prove that $\inf\sigma_{two}(H)= m.$ Let
the conditions of Lemma \ref{inf} be fulfilled. Since the function
$\Delta(0,\cdot)$ is decreasing on $(-\infty,m)$ and the function
$u(\cdot)$ (resp. $\Lambda(\cdot)$) has a unique minimum (resp.
maximum) at $p=0$ for all $z<m$ and $p\in \T^3$ we have
\begin{equation}\label{positivity}
\Delta(p,z)=u(p)-z-\frac{1}{2}\Lambda(p,z)>u(0)-m-\frac{1}{2}\Lambda(0,m).
\end{equation}

If the operator $h(0)$ has either a $m$ energy resonance or the
number $z=m$ is an eigenvalue, then by Lemmas \ref{h0 resonans} and
\ref{zeroeigen} we have  $\Delta(0,m)=0.$ Hence by inequality
\eqref{positivity} we conclude that $\Delta(p,z)>0$ for all $p\in
\T^3$ and $z<m.$ By Lemma \ref{delta=0} the operator $h(p),\,p\in
\T^3$ has no eigenvalues lying  on l.h.s. of $ $ $m.$ Thus,
 $\inf\sigma_{two}(H)= m.$
\end{proof}

\begin{lemma}\label{U.ineq} Let Assumption \ref{hypoth1} be
fulfilled. Then there exist numbers $C_1, C_2,C_3>0$ and $\delta>0$
such that the following inequalities
\begin{align*}
&(i)\quad C_1 (|p|^2+|q|^2)\leq w(p,q)-m \leq C_2 (|p|^2+|q|^2)\quad
\mbox{for all} \quad p, q\in U_\delta(0),\\
 &(ii)\quad w(p,q)-m \geq C_3\quad \mbox{for all} \quad(p, q)\notin
U_\delta(0)\times U_\delta(0)
\end{align*}
 hold.
\end{lemma}
\begin{proof}
 By Assumption \ref{hypoth1} the point $(0,0)\in (\T^3)^2$ is
the unique non-degenerated minimum of the function $w(\cdot,\cdot).$
Then by \eqref{asymp1} there exist positive  numbers $C_1,C_2,C_3$
and a $\delta-$neighborhood of $p=0\in \T^3$ so that $(i)$ and
$(ii)$ hold true.
\end{proof}

 \begin{lemma}\label{razlojeniya-1}
 The right-hand
derivative of $D_1(\cdot)$ at $\zeta=0$ exists and the following
equality
\begin{equation}\label{part D0}
\frac{\partial}{\partial \zeta}D_1(0)=2\sqrt{2}\pi^2
l_1^{-\frac{3}{2}} v^2(0)
  (det W)^{-\frac{1}{2}}
\end{equation}
holds.
\end{lemma}

\begin{proof}
We consider
\begin{align}\label{D1}
D_1(\zeta)-D_1(0)=-\frac{\zeta^2}{2}\int\limits_{U_\delta(0)}
\frac{v^2(q)dq} {(w_0(0,q)+\zeta^2)w_0(0,q)}.
\end{align}

The function  $w_0(0,\cdot)$ has a unique non-degenerate minimum at
$q=0.$ Therefore,  by virtue of the Morse lemma (see \cite{Fed})
there exists a one-to-one mapping $q=\varphi(t)$ of a certain ball
$W_\gamma(0)$  of radius $\gamma>0$ with the center at the origin to
$U_\delta(0)$ such that:
\begin{align}\label{eps=t2}
w_0(0,\varphi(t))=t^2
\end{align}
with $\varphi(0)=0$ and for the Jacobian $J_\varphi(t) \in
\cB(\theta,U_\delta(0))$ of the mapping $q=\varphi(t)$ the equality
$$
J_\varphi(0)=\sqrt{2} l_1^{-\frac{3}{2}}(det W)^{-\frac{1}{2}}
$$
holds, where $\cB(\theta,U_\delta(0))$ can be defined similarly to
$\cB(\theta,\T^3).$

In the integral in \eqref{D1} making a change of variable
$q=\varphi(t)$ and using the equality \eqref{eps=t2} we obtain

\begin{align}\label{D1-D1}
D_1(\zeta)-D_1(0)=-\frac{\zeta^2}{2}\int\limits_{W_\gamma(0)}
\frac{v^2(\varphi(t))J_\varphi(t)}{t^2(t^2+\zeta^2)}dt.
\end{align}

Going over in the integral in \eqref{D1-D1} to spherical coordinates
$t=r\omega,$ we reduce it to the form
\begin{equation}\label{I2}
D_1(\zeta)-D_1(0)=-\frac{\zeta^2}{2}
\int_0^{\gamma}\frac{F(r)}{r^2+\zeta^2} dr,
\end{equation}
with
$$
F(r)=\int_{\S^2}v^2(\varphi(r\omega))J_\varphi(r\omega)d\omega,
$$
where $\S^2$ is the unit sphere in $\R^3$ and $d \omega$ is the
element of the unit sphere in this space.

Using $v,J_\varphi\in \cB(\theta,U_\delta(0))$ we see that
\begin{equation}\label{GEld}
|F(r)-  F(0)|\leq C r^{\theta}.
\end{equation}

Indeed. Since $v,J_\varphi \in \cB(\theta,U_\delta(0))$ for any
$p\in U_\delta(0)$ there exists $C>0$ such that
$$
|v(p)-v(0)|\leq C |p|^\theta\quad \mbox{and} \quad
|J_\varphi(p)-J_\varphi(0)|\leq C |p|^\theta.
$$
Thus
\begin{equation*}
|F(r)-  F(0)|\leq \int\limits_{\S^2}|v^2(\varphi(r\omega))-v^2(0)|
|J_\varphi(r\omega)|d\omega+ \int\limits_{\S^2}|v^2(0)|
|J_\varphi(r\omega)-J_\varphi(0)|d\omega \leq C r^{\theta}.
\end{equation*}

The function $D_1(\zeta)-D_1(0)$ can be rewritten in the form
\begin{equation}\label{F(0)}
D_1(\zeta)-D_1(0)=2\sqrt{2}\pi l_1^{-\frac{3}{2}} v^2(0)
  (det W)^{-\frac{1}{2}} \int_0^\gamma\frac{\zeta^2
dr}{r^2+\zeta^2}+
\end{equation}
$$
\frac{v^2(0)}{2}\int_0^\gamma\frac{\zeta^2(F(r)-F(0))}{r^2+\zeta^2}dr.
$$

By inequality \eqref{GEld} we have
\begin{equation}\label{F-F}
\int_0^\gamma\frac{|F(r)-F(0)|}{r^2+\zeta^2}dr\leq C
\int_0^\gamma\frac{r^\theta}{r^2+\zeta^2}dr.
\end{equation}

Computing the integrals
\begin{equation*}
\int_{0}^{\gamma}
  \frac{\zeta}{r^2+\zeta^2}dr
\quad\mbox{and}\quad \int_{0}^{\gamma}
  \frac{\zeta r^{\theta}}{r^2+\zeta^2}dr
\end{equation*}
we obtain
\begin{equation}\label{Limit}
\int_{0}^{\gamma}
  \frac{\zeta}{r^2+\zeta^2}dr \to \frac{\pi}{2}\,\,\,as\,\,\zeta\to 0+
\quad\mbox{and}\quad \int_{0}^{\gamma}
  \frac{\zeta r^{\theta}}{r^2+\zeta^2}dr\to 0 \,\,\,as\,\,\zeta \to 0+.
\end{equation}

Hence by the equality \eqref{F(0)} and the inequality \eqref{F-F} we
have that there  exists a right-hand derivative of $D_1(\cdot)$ at
$\zeta=0$ and the equality \eqref{part D0} holds.
\end{proof}

{\bf Acknowledgement}

The authors would like to thank Prof.~R.~A.~Minlos and Prof.~H.~
Spohn and Dr.Z.~I.~Muminov for most stimulating discussions on the
results of the paper. This work was supported by DFG 436 USB 113/4
and DFG 436 USB 113/6 projects and the Fundamental Science
Foundation of Uzbekistan. The last two named authors gratefully
acknowledge the hospitality of the Institute of Applied Mathematics
and of the IZKS of the University of Bonn.

\end{document}